\documentclass[letterpaper,twocolumn,english,superscriptaddress,letterpaper, nofootinbib]{revtex4-2}

\usepackage[]{fontenc}
\usepackage[latin9]{inputenc}
\usepackage{textcomp}
\usepackage{amsmath}
\usepackage{graphicx}
\usepackage{amssymb}
\usepackage{bbm}
\usepackage{dsfont}
\usepackage{newtxmath}
\usepackage{dcolumn}
\usepackage{gensymb}
\usepackage{babel}
\usepackage{xcolor}
\usepackage[normalem]{ulem}
\makeatletter

\pdfpageheight\paperheight
\pdfpagewidth\paperwidth

\newcommand{\lyxmathsym}[1]{\ifmmode\begingroup\def\b@ld{bold}
  \text{\ifx\math@version\b@ld\bfseries\fi#1}\endgroup\else#1\fi}

\@ifundefined{textcolor}{}
{
\definecolor{BLACK}{gray}{0}
\definecolor{WHITE}{gray}{1}
\definecolor{RED}{rgb}{1,0,0}
\definecolor{GREEN}{rgb}{0,1,0}
\definecolor{BLUE}{rgb}{0,0,1}
\definecolor{CYAN}{cmyk}{1,0,0,0}
\definecolor{MAGENTA}{cmyk}{0,1,0,0}
\definecolor{YELLOW}{cmyk}{0,0,1,0}
}

\newcommand{\ys}{YbSb$_{2}$}

\definecolor{dark-red}{rgb}{0.9,0.15,0.15}
\definecolor{dark-blue}{rgb}{0.15,0.15,0.4}
\definecolor{medium-blue}{rgb}{0,0,0.5}

\begin{document}
	
\title{Direction-Selective Anomalous Quantum Transport from Saddle-like Topological Surface States Coupled to an Electronic Instability in YbSb$_{2}$}

\author{Rudra Dhara}
\affiliation{Department of Condensed Matter Physics and Material Science,\\ Tata Institute of Fundamental Research, Mumbai, MH 400005, India}
\author{Pritesh Srivastava}
\affiliation{Department of Condensed Matter Physics and Material Science,\\ Tata Institute of Fundamental Research, Mumbai, MH 400005, India}
\author{Jack Tregidga}
\affiliation{Materials Department, University of California, Santa Barbara, California 93106, USA}
\author{Rahul Sharma}
\affiliation{Indian Institute of Science Education and Research Bhopal, Bhopal 462066, India}
\author{Shivani}
\affiliation{Indian Institute of Science Education and Research Bhopal, Bhopal 462066, India}
\author{Shankhadip Bhattacharjee}
\affiliation{Department of Condensed Matter Physics and Material Science,\\ Tata Institute of Fundamental Research, Mumbai, MH 400005, India}
\author{Danilo Ratkovski}
\affiliation{National High Magnetic Field Laboratory, 1800 East Paul Dirac Drive, Tallahassee, Florida 32310-3706, USA}
\author{Shikhar Gupta}
\affiliation{Department of Condensed Matter Physics and Material Science,\\ Tata Institute of Fundamental Research, Mumbai, MH 400005, India}
\author{Kartick Biswas}
\affiliation{Centre for Nano Science and Engineering, Indian Institute of Science, Bengaluru 560012, India}
\author{Shivesh Yadav}
\affiliation{Department of Condensed Matter Physics and Material Science,\\ Tata Institute of Fundamental Research, Mumbai, MH 400005, India}
\author{Nilesh Kulkarni}
\affiliation{Department of Condensed Matter Physics and Material Science,\\ Tata Institute of Fundamental Research, Mumbai, MH 400005, India}
\author{Sarthak S. Gandhi}
\affiliation{Department of Condensed Matter Physics and Material Science,\\ Tata Institute of Fundamental Research, Mumbai, MH 400005, India}
\author{Pavan Nukala}
\affiliation{Centre for Nano Science and Engineering, Indian Institute of Science, Bengaluru 560012, India}
\author{Surajit Saha}
\affiliation{Indian Institute of Science Education and Research Bhopal, Bhopal 462066, India}
\author{John W. Harter}
\affiliation{Materials Department, University of California, Santa Barbara, California 93106, USA}
\author{Bahadur Singh}
\affiliation{Department of Condensed Matter Physics and Material Science,\\ Tata Institute of Fundamental Research, Mumbai, MH 400005, India}
\author{Shouvik Chatterjee}
\email[Author to whom correspondence should be addressed: ]{shouvik.chatterjee@tifr.res.in}
\affiliation{Department of Condensed Matter Physics and Material Science,\\ Tata Institute of Fundamental Research, Mumbai, MH 400005, India}

\begin{abstract}
We report direction-selective anomalous quantum transport in epitaxial \ys\/ thin films associated with an interplay between anisotropic topological surface states and an electronic instability. Quantum oscillations reveal a quasi-two-dimensional orbit with a non-trivial Berry phase and Dirac-like dispersion, observable for current along the \(a\)-axis but absent along the orthogonal in-plane \(c\)-axis. Together with first-principles calculations, these observations support its assignment to a saddle-like topological surface state. Its oscillation amplitude exhibits a field-dependent, non-monotonic temperature evolution that departs from conventional Lifshitz-Kosevich behavior. Raman and coherent phonon spectroscopy reveal additional low-temperature modes consistent with a density-wave instability. Their pronounced polarization anisotropy and colossal field-induced Raman enhancement suggest a strongly anisotropic, field-enhanced electronic susceptibility associated with the density-wave modes. Our findings point to dynamical coupling between anisotropic topological surface states and a field-responsive density-wave instability as a route to direction-selective anomalous quantum transport.

\end{abstract}

\maketitle     
  
Semiclassical Boltzmann theory provides the conventional framework for magnetotransport in metals~\cite{AshcroftMermin1976,Ziman1960}, while quantum oscillations (QO) provide a remarkably precise probe of the underlying electronic Fermi-surface topology, quasiparticle masses, and scattering. The temperature and field dependence of QO amplitude is described by Lifshitz-Kosevich (LK) theory~\cite{Shoenberg1984}.
Deviations from this framework have nevertheless been observed in several quantum materials~\cite{mccollam2005anomalous,shishido2018anomalous,harrison2018highly,wu2019anomalous,galeski2024quantum}, with proposed explanations ranging from quasiparticle renormalization in strongly correlated systems to hydrodynamic transport in Dirac metals. Dynamical coupling to a field-enhanced electronic instability offers another intriguing route to unconventional QO behavior~\cite{trescher2018quantum,mangeolle2025anomalous,mangeolle2026dynamical}, particularly in highly anisotropic topological surface states (TSS), where it may result in strongly direction-dependent magnetotransport.

Here we identify density-wave order accompanying direction-selective anomalous quantum transport in epitaxial \ys\/ thin films. Quantum oscillations reveal a quasi-two-dimensional mode associated with a saddle-like TSS~\cite{singh2018saddle,ghosh2019saddle}, detectable in \(a\)-axis transport but absent along the orthogonal \(c\)-axis. Its amplitude exhibits a pronounced, field-dependent, non-monotonic temperature dependence that departs from conventional LK behavior. Over a similar field and temperature range, \(a\)-axis transport develops a resistivity upturn and butterfly-shaped angular magnetoresistance (AMR), accompanied by progressive suppression of a bulk oscillation. These features are absent in \(c\)-axis transport, whose quantum oscillations exhibit conventional LK behavior. Near 120~K, a resistivity hump, substantially stronger along \(a\)- than \(c\)-axis~\cite{pariari2021interplay}, coincides with the emergence of additional Raman and coherent phonon modes beyond those allowed for the parent \(Cmcm\) structure. The coherent oscillations exhibit pronounced polarization anisotropy, with a larger response along \(a\), while the additional Raman modes display a colossal field-induced intensity enhancement relative to the parent \(A_g\) modes. Together, these observations identify a density-wave state with a strongly anisotropic, field-enhanced electron-lattice response and support its coupling to the topological transport channel as a mechanism for unconventional quantum oscillations and direction-selective magnetotransport.

\ys\/ belongs to the layered square-net family, where non-symmorphic crystal symmetries stabilize topological band crossings~\cite{lee2021topological,klemenz2019topological}. Its orthorhombic $Cmcm$ structure comprises two-dimensional Sb square-nets separated by Yb and Sb zigzag chains~\cite{dhara2025linear}. \ys\/ is characterized by a filling-dependent non-trivial $Z_{2}$ index and is, therefore, expected to host TSS~\cite{kataria2026observation,suppl}. However, the nature of these states and their experimental signatures remain largely unexplored. In addition, a superconducting ground state below $T_{c}\approx 1$ K has been established in both bulk single crystals~\cite{zhao2012type} and thin films~\cite{dhara2025linear}, with a recent study suggesting that the superconductivity may be of unconventional spin-triplet character~\cite{kataria2026observation}.

\begin{figure}
    \centering
    \includegraphics[width=1\linewidth]{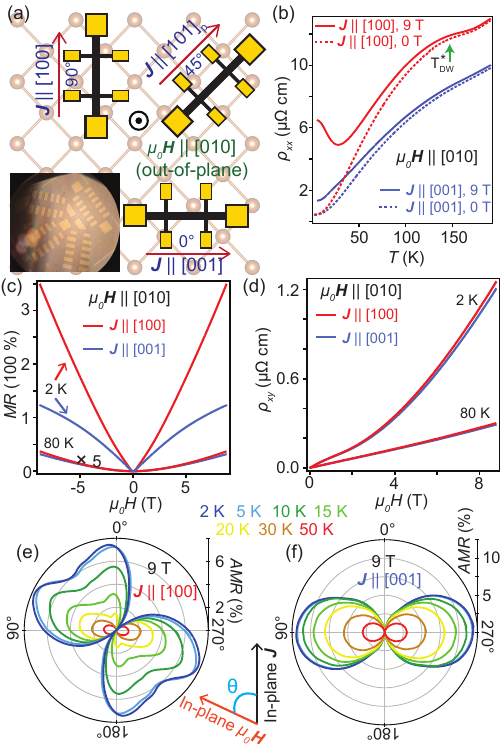}
    \caption{\textbf{Current-direction-dependent anisotropic magnetotransport.} (a) Schematic illustration of the measurement geometry for the directional Hall bar devices. The inset shows an optical microscope image of the fabricated device. (b) Temperature dependence of longitudinal resistivity. Magnetic field dependence of (c) magnetoresistance and (d) Hall resistivity for $\mathbf{J} \parallel [100]$ and $\mathbf{J} \parallel [001]$ with $\mathbf{B} \parallel [010]$. The green arrow in (b) shows the hump-like feature at $T^{*}_{DW} \approx 120~K$. Temperature dependence of AMR at 9~T field for (e) $\mathbf{J} \parallel [100]$ and (f) $\mathbf{J} \parallel [001]$. }

    \label{fig:fig1_Transport}
\end{figure}

Epitaxial thin films of \ys\/ were synthesized on GaSb (001) substrates with the long $b$ axis ($b=16.667$ \AA\/) oriented out of plane and the $a$ ($4.524$ \AA\/) and $c$ ($4.308$ \AA\/) axes lying in the film plane. The epitaxial relationship is (010)[100]\ys\/$\parallel$(001)[110]GaSb, such that the two-dimensional Sb square-net layers of \ys\/ lie in the $ac$ plane. Details of the growth and structural characterization can be found in Ref.~\cite{dhara2025linear} and the Supplemental Material \cite{suppl}. Despite the relatively small in-plane orthorhombicity ($\sim5\%$) in thin films, comparable to that reported for bulk single crystals, we observe pronounced current-direction-dependent magnetotransport in the longitudinal resistivity, Shubnikov-de Haas (SdH) oscillations, and AMR. To systematically probe this anisotropy, we fabricated Hall-bar devices precisely aligned along different crystallographic directions within the $ac$ plane, as shown in Fig.~1(a).

Magnetotransport measurements were performed using a low-frequency ac lock-in technique in a commercial PPMS up to 14~T and at the National High Magnetic Field Laboratory (NHMFL), Tallahassee, up to 35~T. Figures~1(b)-1(d) present the longitudinal resistivity, magnetoresistance (MR), and Hall response for $\mathbf{J}\parallel[100]$ and $[001]$, with $\mathbf{B}\parallel[010]$. At zero field, both directions exhibit metallic behavior, with nearly identical resistivities at 2~K: $\rho_{aa}=0.45~\mu\Omega\mathrm{cm}$ and $\rho_{cc}=0.43~\mu\Omega\mathrm{cm}$. A hump-like feature is observed near 120~K, which is much more pronounced in $\rho_{aa}$, suggestive of the density-wave instability~\cite{pariari2021interplay} discussed below. Increasing field induces a low-temperature upturn in $\rho_{aa}$ that shifts to higher temperatures with magnetic field and eventually saturates, while $\rho_{cc}$ remains metallic \cite{suppl}. The MR is approximately quadratic for both current directions at high temperatures but becomes strongly anisotropic upon cooling, evolving toward quasi-linear, nonsaturating behavior for $\mathbf{J}\parallel[100]$ and sub-linear near-saturation for $\mathbf{J}\parallel[001]$.

This directional contrast also appears in the angular magnetoresistance (AMR), $\mathrm{AMR}\%(\theta)=100\times[\rho(\theta)-\rho_{\min}]/\rho_{\min}$, measured with $\mathbf{B}$ rotating in the $ac$ plane, where $\theta=0^\circ$ corresponds to $\mathbf{B}\parallel\mathbf{J}$. Figures~1(e) and 1(f) show the AMR at 9~T magnetic field for $\mathbf{J}\parallel[100]$ and $[001]$, respectively. Both configurations exhibit twofold patterns at high temperatures and low fields, consistent with orbital MR. This pattern persists for $\mathbf{J}\parallel[001]$ but develops a pronounced butterfly shape at low temperatures and high fields for $\mathbf{J}\parallel[100]$ \cite{suppl}. The observed anomalies in resistivity, MR, and AMR emerge on a common temperature scale near 25~K, suggesting a shared microscopic origin. The Hall response, however, remains nearly identical for the two current directions throughout the investigated temperature and field ranges, consistent with Onsager-Casimir reciprocity in the absence of spontaneous time-reversal-symmetry breaking \cite{onsager1931reciprocal1,onsager1931reciprocal2,casimir1945onsager}.

\begin{figure*}
    \centering
    \includegraphics[width=1\linewidth]{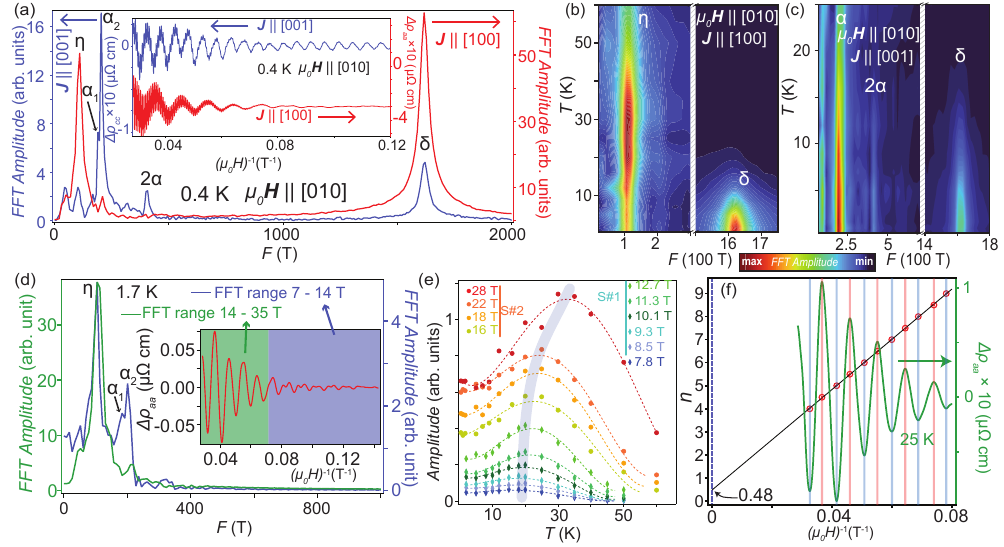}
    \caption{\textbf{Direction-selective anomalous quantum oscillations and $\pi$ Berry phase.}
    (a) FFT spectra of the SdH oscillations at 0.4~K for $\mathbf{J}\parallel[100]$ and $\mathbf{J}\parallel[001]$. The inset shows the $1/B$ dependence of the corresponding background-subtracted oscillatory resistivity ($\Delta\rho_{\mathrm{osc}}$). Temperature evolution of the FFT spectra for (b) $\mathbf{J}\parallel[100]$ and (c) $\mathbf{J}\parallel[001]$.
    (d) FFT spectra obtained over 7--14~T and 14--35~T, showing suppression of the $\alpha$ mode at higher fields for $\mathbf{J}\parallel[100]$. The inset indicates the corresponding FFT windows. A low-pass filter has been used to exclude the $\delta = 1620~T$ mode for clear visualization. (e) Temperature dependence of the oscillation amplitude of the $\eta$ mode at different fields. Dashed lines are guides to the eye. The grey line highlights the field-dependence of the amplitude maxima. S\#1 and S\#2 represent \ys\/ samples with thicknesses of 23~nm and 10~nm, respectively~\cite{suppl}. (f) Landau fan plot of the $\eta$ mode showing a $y$-axis intercept consistent with a $\pi$ Berry phase.}
    \label{fig:Fig2_SdHcomp}
\end{figure*}

The SdH oscillations and their temperature dependence also exhibit pronounced current-direction selectivity with $\mathbf{B}\parallel[010]$ (Fig.~2). For $\mathbf{J}\parallel[001]$, we resolve an $\alpha$ doublet at $\alpha_1=190$~T and $\alpha_2=205$~T and a $\delta$ mode at 1620~T (Fig.~2(a)). Their amplitudes follow the conventional LK temperature dependence (Figs.~2(b),2(c)), yielding effective masses $m_\alpha=0.16m_e$ and $m_\delta=0.58m_e$~\cite{suppl}. Consistent with earlier reports, the $\alpha$ doublet originates from a corrugated zone-edge Fermi surface, whereas $\delta$ is attributed to a high-field magnetic-breakdown orbit~\cite{sato1999effect,gao2026quantum,suppl}. For $\mathbf{J}\parallel[100]$, the $\alpha$ oscillations are progressively suppressed with increasing field, while a new $\eta$ mode at $\approx$ 100~T strengthens (Fig.~2(d)). This mode is absent for $\mathbf{J}\parallel[001]$ and has not been reported in bulk SdH or de Haas-van Alphen (dHvA) measurements~\cite{sato1999effect,gao2026quantum}. Its amplitude exhibits a pronounced departure from LK behavior, reaching a maximum at a field-dependent intermediate temperature (Fig.~2(e)), comparable to the onset scale of the magnetotransport anomalies discussed above.

\begin{figure*}
    \centering
    \includegraphics[width=1\linewidth]{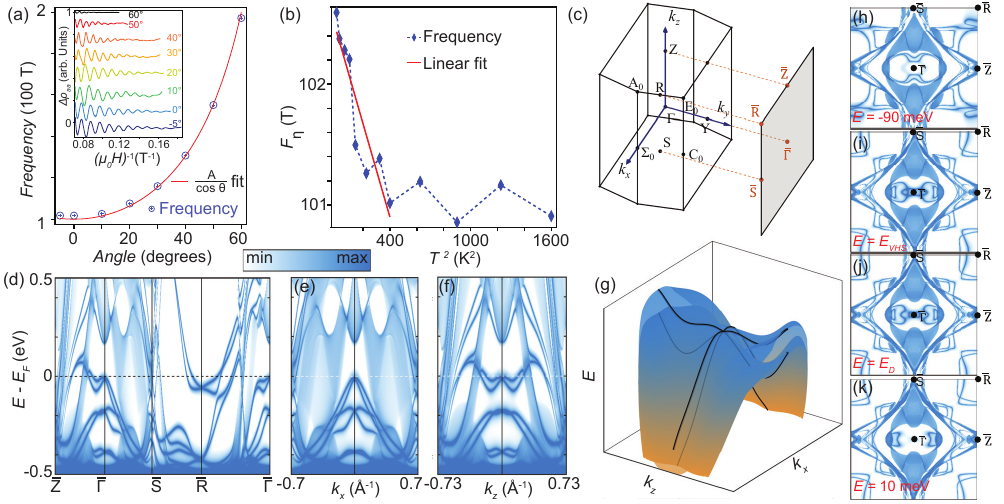}
    \caption {\textbf{Saddle-like topological surface states in YbSb$_{2}$.} Frequency of the $\eta$ orbit as a function of (a) angle between magnetic field and surface normal and (b) $T^{2}$. The red lines in (a) and (b) are fits illustrating $1/cos(\theta)$ and low-temperature linear dependence of the frequency shift, respectively. The inset in (a) shows $\Delta \rho_{osc}$ at different angles. (c) Bulk and (010) surface-projected Brillouin zones of \ys\/. Corresponding high symmetry points are marked in black and red, respectively. (d) (010) surface spectrum in the full surface Brillouin zone. Energy dispersions of the topological surface states along (e) $k_x$ ($\Gamma$-S/[100]) and (f) $k_z$ ($\Gamma$-Z/[001]) directions. (g) Schematic representation of the saddle-like topological surface states. (h-k) Constant-energy contours showing the evolution of the surface states with energy.} 
    \label{TSS}
\end{figure*}

A Landau-fan analysis of the \(\eta\) mode yields a phase consistent with a \(\pi\) Berry phase (Fig.~2(f)), supporting a Dirac-like orbit~\cite{suppl,zhang2005experimental,zhao2022berry,alexandradinata2023fermiology}. Its frequency follows an approximate \(1/\cos\theta\) dependence as the field is tilted from the surface normal, indicating quasi-two-dimensional character (Fig.~3(a))~\cite{Shoenberg1984}. An approximately \(T^2\) frequency shift at low temperatures further supports a Dirac-like dispersion (Fig.~3(b))~\cite{guo2021temperature,alexandradinata2023fermiology}. First-principles calculations reveal a filling-dependent non-trivial \(Z_2\) index~\cite{suppl,kataria2026observation} and saddle-like TSS (Figs.~3(d-g))~\cite{singh2018saddle,ghosh2019saddle}. The absence of rotational symmetry $C_n$ with $n>2$ at the (010) surface of \ys\/ permits saddle-like dispersion, with quasi-Dirac-like behavior along [100] and relatively flat dispersion along [001]. This anisotropy leads to strongly direction-dependent velocities and effective masses, along with generic-$k$, type-II Van Hove singularities (VHSs). The quasi-one-dimensional constant-energy contours change their elongation from [001] at $E-E_F=-90$ meV to [100] at $+10$ meV, passing through the VHSs and the surface Dirac point (Figs. 3(h--k)).
Together, the measurements and calculations support assigning \(\eta\) to a saddle-like TSS, whose anisotropic transport weight naturally explains its detection for \(\mathbf{J}\parallel[100]\) and absence for \(\mathbf{J}\parallel[001]\), with a dispersion consistent with the observed Dirac-like signatures.

Recent theory shows that departures from conventional LK behavior, such as those observed for the $\eta$ orbit, can arise from dynamical coupling of electronic states to fluctuating bosonic modes associated with a density-wave-like instability~\cite{trescher2018quantum,mangeolle2025anomalous,mangeolle2026dynamical}. Through electron-phonon coupling, these correlations can modify lattice dynamics, motivating coherent phonon and Raman spectroscopy as complementary probes of density-wave order. Both measurements employ normal or near-normal incidence along the out-of-plane $b$ axis, with optical polarizations confined to the $ac$ plane. For centrosymmetric \ys\/ ($Cmcm$ (No.~63), point group $mmm$), the zone-center optical modes decompose as
\begin{equation}
\Gamma_{\mathrm{opt}}=3A_g+3B_{1g}+3B_{3g}+2B_{1u}+2B_{2u}+2B_{3u}.
\end{equation}
In coherent phonon spectroscopy, the conventional displacive excitation of coherent phonons (DECP) mechanism excites only fully symmetric modes, restricting the expected oscillatory response to the three $A_g$ phonons. In Raman scattering, the six infrared-active modes are forbidden by inversion symmetry, while the $B_{1g}$ and $B_{3g}$ tensors require a polarization component along $b$ and are therefore inaccessible in our normal-incidence geometry. Consistent with these selection rules, both measurements resolve three high-temperature $A_g$ modes, labeled $P_1$, $P_2$, and $P_3$ (Figs.~4(b),4(d)), with frequencies in good agreement with first-principles calculations~\cite{suppl}.

\begin{figure*}
    \centering
    \includegraphics[width=1\linewidth]{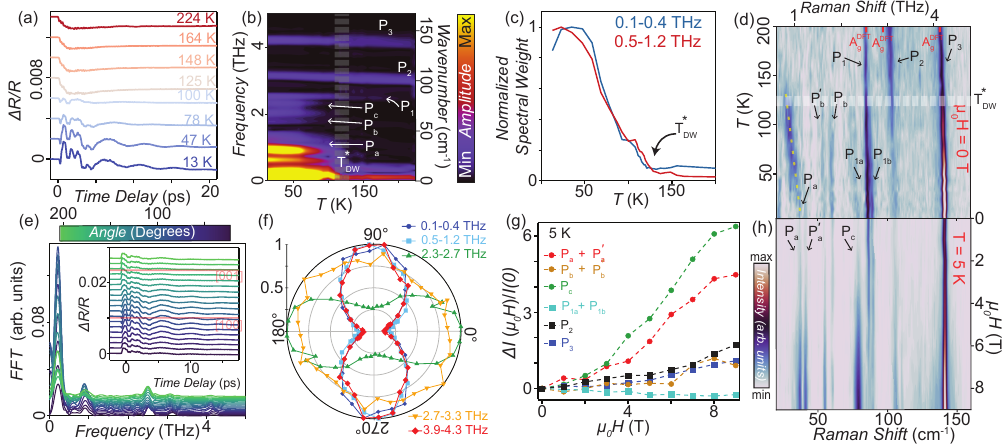}
    \caption{\textbf{Coherent phonon and Raman spectroscopy of YbSb$_{2}$.}~(a) Transient changes in reflectivity ($\Delta R/R$) as a function of pump-probe time delay for selected temperatures. Curves are offset for clarity. (b) Temperature-frequency map of the corresponding coherent oscillations induced by the pump pulse, extracted through Fourier transforms. (c) Temperature dependence of the integrated spectral weight of the low-frequency modes between 0.1-0.4~THz and 0.5-1.2~THz. The finite spectral weight at high temperatures is likely due to incomplete background subtraction. (d) Temperature-frequency map of the Raman modes at zero magnetic field. The dotted line highlights the strong mode softening of the P$_{a}$ mode. (e) Angular dependence of the coherent oscillations, where the sample was rotated about its surface normal in the $ac$ plane keeping the pump and probe polarizations fixed. Curves are offset for clarity. The inset shows the corresponding angular dependence of the transient reflectivity traces. (f) Polar plots showing the angular dependence of the integrated spectral weight of the coherent oscillations. $0^\circ$ corresponds to polarization parallel to [001] for data in (e) and (f). (g) Relative change in the integrated Raman spectral weight, \(\Delta I(B)/I(0)=(I(B)-I(0))/I(0)\), for the indicated modes at 5~K. Dashed lines are guides to the eye. (h) Magnetic field-frequency map of the Raman modes at 5~K.}
\end{figure*}

Upon cooling below \(T^{*}_{\mathrm{DW}}\approx120\)~K, coherent phonon and Raman spectra develop additional modes beyond those allowed for the parent \(Cmcm\) structure in our geometry (Figs.~4(a),4(b),4(d)). The integrated spectral weights of the additional coherent phonon modes exhibit order-parameter-like behavior (Fig.~4(c)), with an onset near $T^{*}_{DW}$, where the resistivity shows a hump-like feature (Fig.~1(b)). These observations support a density-wave instability that breaks translational symmetry, folds finite-wave-vector phonons to the zone center, and activates additional modes. The low-frequency modes at 0.2~THz and 0.9~THz (\(P_a\)) soften and become increasingly damped on warming toward \(T^{*}_{\mathrm{DW}}\) (Figs.~4(b),4(d)), suggesting their assignment as density-wave amplitude modes. Both exhibit pronounced twofold intensity anisotropy as the sample is rotated relative to the fixed optical polarizations within the \(ac\) plane (Figs.~4(e),4(f)). Among the parent \(A_g\) modes, only \(P_3\), which arises from vibrations of the Sb square-net atoms~\cite{suppl}, exhibits comparably strong anisotropy (Fig.~4(f)). This correspondence suggests a prominent role for the Sb square net, consistent with charge-density-wave (CDW) instabilities involving Sb-derived electronic states in related compounds, including LaAuSb\(_2\) and LaSb\(_2\)~\cite{wu2023coexistence,palacio2023fermi}. Although permitted by the orthorhombic symmetry, the pronounced polarization dependence supports an anisotropic electronic response involving the Sb square net, consistent with the directional magnetotransport anomalies.

Magnetic field induces a colossal enhancement of select additional low-temperature Raman modes ($P_a$, $P_c$) in nonmagnetic \ys\/ (Figs.~4(g),4(h)), exceeding the giant magneto-optical response reported in MoS\(_2\)~\cite{ji2016giant} and resembling the enhancement of axial-Higgs CDW modes in GdTe\(_3\)~\cite{wulferding2025magnetic}. The parent \(A_g\) modes exhibit smaller, contrasting changes: \(P_{1a}\) and \(P_{1b}\) lose spectral weight, whereas \(P_2\) and \(P_3\) gain intensity (Figs.~4(g),4(h)). This mode-selective redistribution suggests an anisotropic, field-dependent electronic susceptibility coupled to the lattice. The strong field response of the density-wave-associated modes supports a scenario in which dynamical coupling to these excitations produces the observed departures from conventional LK behavior~\cite{trescher2018quantum,mangeolle2025anomalous,mangeolle2026dynamical}.

Our results provide converging evidence for a field-responsive density-wave instability coupled to anisotropic electronic states in \ys\/, offering a microscopic framework for direction-selective magnetotransport and unconventional quantum oscillations. Determining the ordering wave vector, microscopic distortion, and broken symmetries remains an important direction for future work. The resemblance to the field response of axial-Higgs CDW modes motivates further symmetry-sensitive measurements, including tests of time-reversal symmetry, particularly given recent \(\mu\)SR evidence for its breaking in the superconducting state~\cite{kataria2026observation}. The presence of saddle-like TSSs with Type-II VHSs may drive surface superconducting pairing, potentially favoring topological superconductivity ~\cite{yao2015topological,meng2015evidence}. To our knowledge, neither density-wave order nor the associated resistivity hump and pronounced directional transport anomalies have been reported in bulk \ys\/. More broadly, understanding and controlling the competition or cooperation between density-wave order and superconductivity remains an outstanding challenge across unconventional superconductors. Our findings suggest that epitaxial engineering of square-net materials offers a promising route to tuning the balance between these intertwined orders.


\section*{Acknowledgments}

We thank Devendra Buddhikot, Ruta Kulkarni, and Ganesh Jangam for technical assistance. We acknowledge the Anusandhan National Research Foundation (ANRF), Award No. ANRF/ARG/2025/001541/PS, and the Department of Atomic Energy (DAE) of the Government of India, Project Identification Nos. RTI4013 and RTI4015, for support. Coherent phonon spectroscopy experiments were supported by the U.S. National Science Foundation under Award No.~DMR-2140786. SS acknowledges ANRF for funding through Award No. CRG/2023/000581. The authors acknowledge the Micro and Nano Characterization Facility (MNCF), Centre for Nano Science and Engineering (CeNSE), Indian Institute of Science (IISc), Bengaluru, for FIB-based TEM specimen preparation, and the Advanced Facility for Microscopy and Microanalysis (AFMM), IISc, for providing access to the HRSTEM imaging facility. A portion of this work was performed at the National High Magnetic Field Laboratory, which is supported by the U.S. National Science Foundation Cooperative Agreement No.~DMR-2128556 and the State of Florida. 


\section*{Author Contributions}

S.C. and R.D. planned the experiments. R.D., with assistance from S.B. and N.K., fabricated the thin films and performed structural characterization. R.D., with assistance from S.G. and  S.Y., fabricated the Hall bar devices. R.D., with assistance from S.B., S.G., S.Y., and S.S.G., performed the electrical and magnetometry measurements. S.C., with assistance from D.R., performed magnetotransport measurements under high magnetic field at NHMFL. P.S., under the guidance of B.S., performed first-principles calculations. J.T., under the guidance of J.W.H., performed transient reflectivity and coherent phonon spectroscopy measurements. R.S. and S., under the guidance of S.S., performed Raman spectroscopy. K.B., under the guidance of P.N., prepared thin-film lamellae and performed HAADF-STEM measurements. R.D. and S.C. wrote the manuscript with inputs from B.S., J.W.H. and S.S. S.C. conceived the project and was responsible for its overall execution. All authors discussed the results and commented on the manuscript.

\section*{Data Availability}

The data that support the findings of this study are available from the corresponding author upon reasonable request.

\section*{Competing Interests}
The authors declare no competing interests.

\bibliographystyle{apsrev4-1}  
\bibliography{Main.bib}

\begin{thebibliography}{35}%
\makeatletter
\providecommand \@ifxundefined [1]{%
 \@ifx{#1\undefined}
}%
\providecommand \@ifnum [1]{%
 \ifnum #1\expandafter \@firstoftwo
 \else \expandafter \@secondoftwo
 \fi
}%
\providecommand \@ifx [1]{%
 \ifx #1\expandafter \@firstoftwo
 \else \expandafter \@secondoftwo
 \fi
}%
\providecommand \natexlab [1]{#1}%
\providecommand \enquote  [1]{``#1''}%
\providecommand \bibnamefont  [1]{#1}%
\providecommand \bibfnamefont [1]{#1}%
\providecommand \citenamefont [1]{#1}%
\providecommand \href@noop [0]{\@secondoftwo}%
\providecommand \href [0]{\begingroup \@sanitize@url \@href}%
\providecommand \@href[1]{\@@startlink{#1}\@@href}%
\providecommand \@@href[1]{\endgroup#1\@@endlink}%
\providecommand \@sanitize@url [0]{\catcode `\\12\catcode `\$12\catcode
  `\&12\catcode `\#12\catcode `\^12\catcode `\_12\catcode `\%12\relax}%
\providecommand \@@startlink[1]{}%
\providecommand \@@endlink[0]{}%
\providecommand \url  [0]{\begingroup\@sanitize@url \@url }%
\providecommand \@url [1]{\endgroup\@href {#1}{\urlprefix }}%
\providecommand \urlprefix  [0]{URL }%
\providecommand \Eprint [0]{\href }%
\providecommand \doibase [0]{http://dx.doi.org/}%
\providecommand \selectlanguage [0]{\@gobble}%
\providecommand \bibinfo  [0]{\@secondoftwo}%
\providecommand \bibfield  [0]{\@secondoftwo}%
\providecommand \translation [1]{[#1]}%
\providecommand \BibitemOpen [0]{}%
\providecommand \bibitemStop [0]{}%
\providecommand \bibitemNoStop [0]{.\EOS\space}%
\providecommand \EOS [0]{\spacefactor3000\relax}%
\providecommand \BibitemShut  [1]{\csname bibitem#1\endcsname}%
\let\auto@bib@innerbib\@empty
\bibitem [{\citenamefont {Ashcroft}\ and\ \citenamefont
  {Mermin}(1976)}]{AshcroftMermin1976}%
  \BibitemOpen
  \bibfield  {author} {\bibinfo {author} {\bibfnamefont {N.~W.}\ \bibnamefont
  {Ashcroft}}\ and\ \bibinfo {author} {\bibfnamefont {N.~D.}\ \bibnamefont
  {Mermin}},\ }\href@noop {} {\emph {\bibinfo {title} {Solid State Physics}}}\
  (\bibinfo  {publisher} {Holt, Rinehart and Winston},\ \bibinfo {address} {New
  York},\ \bibinfo {year} {1976})\BibitemShut {NoStop}%
\bibitem [{\citenamefont {Ziman}(1960)}]{Ziman1960}%
  \BibitemOpen
  \bibfield  {author} {\bibinfo {author} {\bibfnamefont {J.~M.}\ \bibnamefont
  {Ziman}},\ }\href@noop {} {\emph {\bibinfo {title} {Electrons and Phonons:
  The Theory of Transport Phenomena in Solids}}}\ (\bibinfo  {publisher}
  {Clarendon Press},\ \bibinfo {address} {Oxford},\ \bibinfo {year}
  {1960})\BibitemShut {NoStop}%
\bibitem [{\citenamefont {Shoenberg}(1984)}]{Shoenberg1984}%
  \BibitemOpen
  \bibfield  {author} {\bibinfo {author} {\bibfnamefont {D.}~\bibnamefont
  {Shoenberg}},\ }\href {\doibase 10.1017/CBO9780511897870} {\emph {\bibinfo
  {title} {Magnetic Oscillations in Metals}}}\ (\bibinfo  {publisher}
  {Cambridge University Press},\ \bibinfo {address} {Cambridge},\ \bibinfo
  {year} {1984})\BibitemShut {NoStop}%
\bibitem [{\citenamefont {McCollam}\ \emph {et~al.}(2005)\citenamefont
  {McCollam}, \citenamefont {Julian}, \citenamefont {Rourke}, \citenamefont
  {Aoki},\ and\ \citenamefont {Flouquet}}]{mccollam2005anomalous}%
  \BibitemOpen
  \bibfield  {author} {\bibinfo {author} {\bibfnamefont {A.}~\bibnamefont
  {McCollam}}, \bibinfo {author} {\bibfnamefont {S.~R.}\ \bibnamefont
  {Julian}}, \bibinfo {author} {\bibfnamefont {P.~M.~C.}\ \bibnamefont
  {Rourke}}, \bibinfo {author} {\bibfnamefont {D.}~\bibnamefont {Aoki}}, \ and\
  \bibinfo {author} {\bibfnamefont {J.}~\bibnamefont {Flouquet}},\ }\href
  {\doibase 10.1103/PhysRevLett.94.186401} {\bibfield  {journal} {\bibinfo
  {journal} {Physical Review Letters}\ }\textbf {\bibinfo {volume} {94}},\
  \bibinfo {pages} {186401} (\bibinfo {year} {2005})}\BibitemShut {NoStop}%
\bibitem [{\citenamefont {Shishido}\ \emph {et~al.}(2018)\citenamefont
  {Shishido}, \citenamefont {Yamada}, \citenamefont {Sugii}, \citenamefont
  {Shimozawa}, \citenamefont {Yanase},\ and\ \citenamefont
  {Yamashita}}]{shishido2018anomalous}%
  \BibitemOpen
  \bibfield  {author} {\bibinfo {author} {\bibfnamefont {H.}~\bibnamefont
  {Shishido}}, \bibinfo {author} {\bibfnamefont {S.}~\bibnamefont {Yamada}},
  \bibinfo {author} {\bibfnamefont {K.}~\bibnamefont {Sugii}}, \bibinfo
  {author} {\bibfnamefont {M.}~\bibnamefont {Shimozawa}}, \bibinfo {author}
  {\bibfnamefont {Y.}~\bibnamefont {Yanase}}, \ and\ \bibinfo {author}
  {\bibfnamefont {M.}~\bibnamefont {Yamashita}},\ }\href {\doibase
  10.1103/PhysRevLett.120.177201} {\bibfield  {journal} {\bibinfo  {journal}
  {Physical Review Letters}\ }\textbf {\bibinfo {volume} {120}},\ \bibinfo
  {pages} {177201} (\bibinfo {year} {2018})}\BibitemShut {NoStop}%
\bibitem [{\citenamefont {Harrison}(2018)}]{harrison2018highly}%
  \BibitemOpen
  \bibfield  {author} {\bibinfo {author} {\bibfnamefont {N.}~\bibnamefont
  {Harrison}},\ }\href {\doibase 10.1103/PhysRevLett.121.026602} {\bibfield
  {journal} {\bibinfo  {journal} {Physical Review Letters}\ }\textbf {\bibinfo
  {volume} {121}},\ \bibinfo {pages} {026602} (\bibinfo {year}
  {2018})}\BibitemShut {NoStop}%
\bibitem [{\citenamefont {Wu}\ \emph {et~al.}(2019)\citenamefont {Wu},
  \citenamefont {Guo}, \citenamefont {Smidman}, \citenamefont {Zhang},
  \citenamefont {Chen}, \citenamefont {Singleton},\ and\ \citenamefont
  {Yuan}}]{wu2019anomalous}%
  \BibitemOpen
  \bibfield  {author} {\bibinfo {author} {\bibfnamefont {F.}~\bibnamefont
  {Wu}}, \bibinfo {author} {\bibfnamefont {C.}~\bibnamefont {Guo}}, \bibinfo
  {author} {\bibfnamefont {M.}~\bibnamefont {Smidman}}, \bibinfo {author}
  {\bibfnamefont {J.}~\bibnamefont {Zhang}}, \bibinfo {author} {\bibfnamefont
  {Y.}~\bibnamefont {Chen}}, \bibinfo {author} {\bibfnamefont {J.}~\bibnamefont
  {Singleton}}, \ and\ \bibinfo {author} {\bibfnamefont {H.}~\bibnamefont
  {Yuan}},\ }\href {\doibase 10.1038/s41535-019-0161-4} {\bibfield  {journal}
  {\bibinfo  {journal} {npj Quantum Materials}\ }\textbf {\bibinfo {volume}
  {4}},\ \bibinfo {pages} {20} (\bibinfo {year} {2019})}\BibitemShut {NoStop}%
\bibitem [{\citenamefont {Galeski}\ \emph {et~al.}(2024)\citenamefont
  {Galeski}, \citenamefont {Araki}, \citenamefont {Forslund}, \citenamefont
  {Wawrzy{\'n}czak}, \citenamefont {Legg}, \citenamefont {Sivakumar},
  \citenamefont {Miniotaite}, \citenamefont {Elson}, \citenamefont
  {M{\aa}nsson}, \citenamefont {Witteveen}, \citenamefont {von Rohr},
  \citenamefont {Baron}, \citenamefont {Ishikawa}, \citenamefont {Li},
  \citenamefont {Gu}, \citenamefont {Zhao}, \citenamefont {Zhu}, \citenamefont
  {Chen}, \citenamefont {Wang}, \citenamefont {Parkin}, \citenamefont
  {Grobunov}, \citenamefont {Zherlitsyn}, \citenamefont {Vlaar}, \citenamefont
  {Nguyen}, \citenamefont {Paschen}, \citenamefont {Narang}, \citenamefont
  {Felser}, \citenamefont {Wosnitza}, \citenamefont {Meng}, \citenamefont
  {Sassa}, \citenamefont {Hartnoll},\ and\ \citenamefont
  {Gooth}}]{galeski2024quantum}%
  \BibitemOpen
  \bibfield  {author} {\bibinfo {author} {\bibfnamefont {S.}~\bibnamefont
  {Galeski}}, \bibinfo {author} {\bibfnamefont {K.}~\bibnamefont {Araki}},
  \bibinfo {author} {\bibfnamefont {O.~K.}\ \bibnamefont {Forslund}}, \bibinfo
  {author} {\bibfnamefont {R.}~\bibnamefont {Wawrzy{\'n}czak}}, \bibinfo
  {author} {\bibfnamefont {H.~F.}\ \bibnamefont {Legg}}, \bibinfo {author}
  {\bibfnamefont {P.~K.}\ \bibnamefont {Sivakumar}}, \bibinfo {author}
  {\bibfnamefont {U.}~\bibnamefont {Miniotaite}}, \bibinfo {author}
  {\bibfnamefont {F.}~\bibnamefont {Elson}}, \bibinfo {author} {\bibfnamefont
  {M.}~\bibnamefont {M{\aa}nsson}}, \bibinfo {author} {\bibfnamefont
  {C.}~\bibnamefont {Witteveen}}, \bibinfo {author} {\bibfnamefont {F.~O.}\
  \bibnamefont {von Rohr}}, \bibinfo {author} {\bibfnamefont {A.~Q.~R.}\
  \bibnamefont {Baron}}, \bibinfo {author} {\bibfnamefont {D.}~\bibnamefont
  {Ishikawa}}, \bibinfo {author} {\bibfnamefont {Q.}~\bibnamefont {Li}},
  \bibinfo {author} {\bibfnamefont {G.}~\bibnamefont {Gu}}, \bibinfo {author}
  {\bibfnamefont {L.~X.}\ \bibnamefont {Zhao}}, \bibinfo {author}
  {\bibfnamefont {W.~L.}\ \bibnamefont {Zhu}}, \bibinfo {author} {\bibfnamefont
  {G.~F.}\ \bibnamefont {Chen}}, \bibinfo {author} {\bibfnamefont
  {Y.}~\bibnamefont {Wang}}, \bibinfo {author} {\bibfnamefont {S.~S.~P.}\
  \bibnamefont {Parkin}}, \bibinfo {author} {\bibfnamefont {D.}~\bibnamefont
  {Grobunov}}, \bibinfo {author} {\bibfnamefont {S.}~\bibnamefont
  {Zherlitsyn}}, \bibinfo {author} {\bibfnamefont {B.}~\bibnamefont {Vlaar}},
  \bibinfo {author} {\bibfnamefont {D.~H.}\ \bibnamefont {Nguyen}}, \bibinfo
  {author} {\bibfnamefont {S.}~\bibnamefont {Paschen}}, \bibinfo {author}
  {\bibfnamefont {P.}~\bibnamefont {Narang}}, \bibinfo {author} {\bibfnamefont
  {C.}~\bibnamefont {Felser}}, \bibinfo {author} {\bibfnamefont
  {J.}~\bibnamefont {Wosnitza}}, \bibinfo {author} {\bibfnamefont
  {T.}~\bibnamefont {Meng}}, \bibinfo {author} {\bibfnamefont {Y.}~\bibnamefont
  {Sassa}}, \bibinfo {author} {\bibfnamefont {S.~A.}\ \bibnamefont {Hartnoll}},
  \ and\ \bibinfo {author} {\bibfnamefont {J.}~\bibnamefont {Gooth}},\ }\href
  {\doibase 10.1103/PhysRevB.110.L121103} {\bibfield  {journal} {\bibinfo
  {journal} {Physical Review B}\ }\textbf {\bibinfo {volume} {110}},\ \bibinfo
  {pages} {L121103} (\bibinfo {year} {2024})}\BibitemShut {NoStop}%
\bibitem [{\citenamefont {Trescher}\ \emph {et~al.}(2018)\citenamefont
  {Trescher}, \citenamefont {Bergholtz},\ and\ \citenamefont
  {Knolle}}]{trescher2018quantum}%
  \BibitemOpen
  \bibfield  {author} {\bibinfo {author} {\bibfnamefont {M.}~\bibnamefont
  {Trescher}}, \bibinfo {author} {\bibfnamefont {E.~J.}\ \bibnamefont
  {Bergholtz}}, \ and\ \bibinfo {author} {\bibfnamefont {J.}~\bibnamefont
  {Knolle}},\ }\href {\doibase 10.1103/PhysRevB.98.125304} {\bibfield
  {journal} {\bibinfo  {journal} {Physical Review B}\ }\textbf {\bibinfo
  {volume} {98}},\ \bibinfo {pages} {125304} (\bibinfo {year}
  {2018})}\BibitemShut {NoStop}%
\bibitem [{\citenamefont {Mangeolle}\ and\ \citenamefont
  {Knolle}(2025)}]{mangeolle2025anomalous}%
  \BibitemOpen
  \bibfield  {author} {\bibinfo {author} {\bibfnamefont {L.}~\bibnamefont
  {Mangeolle}}\ and\ \bibinfo {author} {\bibfnamefont {J.}~\bibnamefont
  {Knolle}},\ }\href {\doibase 10.1103/PhysRevLett.134.146502} {\bibfield
  {journal} {\bibinfo  {journal} {Physical Review Letters}\ }\textbf {\bibinfo
  {volume} {134}},\ \bibinfo {pages} {146502} (\bibinfo {year}
  {2025})}\BibitemShut {NoStop}%
\bibitem [{\citenamefont {Mangeolle}\ and\ \citenamefont
  {Knolle}(2026)}]{mangeolle2026dynamical}%
  \BibitemOpen
  \bibfield  {author} {\bibinfo {author} {\bibfnamefont {L.}~\bibnamefont
  {Mangeolle}}\ and\ \bibinfo {author} {\bibfnamefont {J.}~\bibnamefont
  {Knolle}},\ }\href@noop {} {\bibfield  {journal} {\bibinfo  {journal}
  {Physical Review B}\ }\textbf {\bibinfo {volume} {113}},\ \bibinfo {pages}
  {245132} (\bibinfo {year} {2026})}\BibitemShut {NoStop}%
\bibitem [{\citenamefont {Singh}\ \emph {et~al.}(2018)\citenamefont {Singh},
  \citenamefont {Zhou}, \citenamefont {Lin},\ and\ \citenamefont
  {Bansil}}]{singh2018saddle}%
  \BibitemOpen
  \bibfield  {author} {\bibinfo {author} {\bibfnamefont {B.}~\bibnamefont
  {Singh}}, \bibinfo {author} {\bibfnamefont {X.}~\bibnamefont {Zhou}},
  \bibinfo {author} {\bibfnamefont {H.}~\bibnamefont {Lin}}, \ and\ \bibinfo
  {author} {\bibfnamefont {A.}~\bibnamefont {Bansil}},\ }\href {\doibase
  10.1103/PhysRevB.97.075125} {\bibfield  {journal} {\bibinfo  {journal}
  {Physical Review B}\ }\textbf {\bibinfo {volume} {97}},\ \bibinfo {pages}
  {075125} (\bibinfo {year} {2018})}\BibitemShut {NoStop}%
\bibitem [{\citenamefont {Ghosh}\ \emph {et~al.}(2019)\citenamefont {Ghosh},
  \citenamefont {Mardanya}, \citenamefont {Singh}, \citenamefont {Zhou},
  \citenamefont {Wang}, \citenamefont {Chang}, \citenamefont {Su},
  \citenamefont {Lin}, \citenamefont {Agarwal},\ and\ \citenamefont
  {Bansil}}]{ghosh2019saddle}%
  \BibitemOpen
  \bibfield  {author} {\bibinfo {author} {\bibfnamefont {B.}~\bibnamefont
  {Ghosh}}, \bibinfo {author} {\bibfnamefont {S.}~\bibnamefont {Mardanya}},
  \bibinfo {author} {\bibfnamefont {B.}~\bibnamefont {Singh}}, \bibinfo
  {author} {\bibfnamefont {X.}~\bibnamefont {Zhou}}, \bibinfo {author}
  {\bibfnamefont {B.}~\bibnamefont {Wang}}, \bibinfo {author} {\bibfnamefont
  {T.-R.}\ \bibnamefont {Chang}}, \bibinfo {author} {\bibfnamefont
  {C.}~\bibnamefont {Su}}, \bibinfo {author} {\bibfnamefont {H.}~\bibnamefont
  {Lin}}, \bibinfo {author} {\bibfnamefont {A.}~\bibnamefont {Agarwal}}, \ and\
  \bibinfo {author} {\bibfnamefont {A.}~\bibnamefont {Bansil}},\ }\href
  {\doibase 10.1103/PhysRevB.100.235101} {\bibfield  {journal} {\bibinfo
  {journal} {Physical Review B}\ }\textbf {\bibinfo {volume} {100}},\ \bibinfo
  {pages} {235101} (\bibinfo {year} {2019})}\BibitemShut {NoStop}%
\bibitem [{\citenamefont {Pariari}\ \emph {et~al.}(2021)\citenamefont
  {Pariari}, \citenamefont {Koley}, \citenamefont {Roy}, \citenamefont
  {Singha}, \citenamefont {Laad}, \citenamefont {Taraphder},\ and\
  \citenamefont {Mandal}}]{pariari2021interplay}%
  \BibitemOpen
  \bibfield  {author} {\bibinfo {author} {\bibfnamefont {A.}~\bibnamefont
  {Pariari}}, \bibinfo {author} {\bibfnamefont {S.}~\bibnamefont {Koley}},
  \bibinfo {author} {\bibfnamefont {S.}~\bibnamefont {Roy}}, \bibinfo {author}
  {\bibfnamefont {R.}~\bibnamefont {Singha}}, \bibinfo {author} {\bibfnamefont
  {M.~S.}\ \bibnamefont {Laad}}, \bibinfo {author} {\bibfnamefont
  {A.}~\bibnamefont {Taraphder}}, \ and\ \bibinfo {author} {\bibfnamefont
  {P.}~\bibnamefont {Mandal}},\ }\href@noop {} {\bibfield  {journal} {\bibinfo
  {journal} {Physical Review B}\ }\textbf {\bibinfo {volume} {104}},\ \bibinfo
  {pages} {155147} (\bibinfo {year} {2021})}\BibitemShut {NoStop}%
\bibitem [{\citenamefont {Lee}\ \emph {et~al.}(2021)\citenamefont {Lee},
  \citenamefont {Hyun},\ and\ \citenamefont {Shim}}]{lee2021topological}%
  \BibitemOpen
  \bibfield  {author} {\bibinfo {author} {\bibfnamefont {I.}~\bibnamefont
  {Lee}}, \bibinfo {author} {\bibfnamefont {S.}~\bibnamefont {Hyun}}, \ and\
  \bibinfo {author} {\bibfnamefont {J.}~\bibnamefont {Shim}},\ }\href@noop {}
  {\bibfield  {journal} {\bibinfo  {journal} {Physical Review B}\ }\textbf
  {\bibinfo {volume} {103}},\ \bibinfo {pages} {165106} (\bibinfo {year}
  {2021})}\BibitemShut {NoStop}%
\bibitem [{\citenamefont {Klemenz}\ \emph {et~al.}(2019)\citenamefont
  {Klemenz}, \citenamefont {Lei},\ and\ \citenamefont
  {Schoop}}]{klemenz2019topological}%
  \BibitemOpen
  \bibfield  {author} {\bibinfo {author} {\bibfnamefont {S.}~\bibnamefont
  {Klemenz}}, \bibinfo {author} {\bibfnamefont {S.}~\bibnamefont {Lei}}, \ and\
  \bibinfo {author} {\bibfnamefont {L.~M.}\ \bibnamefont {Schoop}},\
  }\href@noop {} {\bibfield  {journal} {\bibinfo  {journal} {Annual review of
  materials research}\ }\textbf {\bibinfo {volume} {49}},\ \bibinfo {pages}
  {185} (\bibinfo {year} {2019})}\BibitemShut {NoStop}%
\bibitem [{\citenamefont {Dhara}\ \emph {et~al.}(2025)\citenamefont {Dhara},
  \citenamefont {Das}, \citenamefont {Dutta}, \citenamefont {Gupta},
  \citenamefont {Yadav}, \citenamefont {Kulkarni}, \citenamefont {Satpati},
  \citenamefont {Raychaudhuri},\ and\ \citenamefont
  {Chatterjee}}]{dhara2025linear}%
  \BibitemOpen
  \bibfield  {author} {\bibinfo {author} {\bibfnamefont {R.}~\bibnamefont
  {Dhara}}, \bibinfo {author} {\bibfnamefont {P.}~\bibnamefont {Das}}, \bibinfo
  {author} {\bibfnamefont {S.}~\bibnamefont {Dutta}}, \bibinfo {author}
  {\bibfnamefont {S.}~\bibnamefont {Gupta}}, \bibinfo {author} {\bibfnamefont
  {S.}~\bibnamefont {Yadav}}, \bibinfo {author} {\bibfnamefont
  {N.}~\bibnamefont {Kulkarni}}, \bibinfo {author} {\bibfnamefont
  {B.}~\bibnamefont {Satpati}}, \bibinfo {author} {\bibfnamefont
  {P.}~\bibnamefont {Raychaudhuri}}, \ and\ \bibinfo {author} {\bibfnamefont
  {S.}~\bibnamefont {Chatterjee}},\ }\href@noop {} {\bibfield  {journal}
  {\bibinfo  {journal} {Physical Review Materials}\ }\textbf {\bibinfo {volume}
  {9}},\ \bibinfo {pages} {034801} (\bibinfo {year} {2025})}\BibitemShut
  {NoStop}%
\bibitem [{\citenamefont {Kataria}\ \emph {et~al.}(2026)\citenamefont
  {Kataria}, \citenamefont {Srivastava}, \citenamefont {Samanta}, \citenamefont
  {Yadav}, \citenamefont {Manna}, \citenamefont {Sharma}, \citenamefont
  {Mishra}, \citenamefont {Barker}, \citenamefont {Hillier}, \citenamefont
  {Agarwal}, \citenamefont {Ghosh},\ and\ \citenamefont
  {Singh}}]{kataria2026observation}%
  \BibitemOpen
  \bibfield  {author} {\bibinfo {author} {\bibfnamefont {A.}~\bibnamefont
  {Kataria}}, \bibinfo {author} {\bibfnamefont {S.}~\bibnamefont {Srivastava}},
  \bibinfo {author} {\bibfnamefont {D.}~\bibnamefont {Samanta}}, \bibinfo
  {author} {\bibfnamefont {P.}~\bibnamefont {Yadav}}, \bibinfo {author}
  {\bibfnamefont {P.}~\bibnamefont {Manna}}, \bibinfo {author} {\bibfnamefont
  {S.}~\bibnamefont {Sharma}}, \bibinfo {author} {\bibfnamefont
  {P.}~\bibnamefont {Mishra}}, \bibinfo {author} {\bibfnamefont
  {J.}~\bibnamefont {Barker}}, \bibinfo {author} {\bibfnamefont {A.~D.}\
  \bibnamefont {Hillier}}, \bibinfo {author} {\bibfnamefont {A.}~\bibnamefont
  {Agarwal}}, \bibinfo {author} {\bibfnamefont {S.~K.}\ \bibnamefont {Ghosh}},
  \ and\ \bibinfo {author} {\bibfnamefont {R.~P.}\ \bibnamefont {Singh}},\
  }\href {\doibase 10.1103/drzq-lfn5} {\bibfield  {journal} {\bibinfo
  {journal} {Physical Review Letters}\ }\textbf {\bibinfo {volume} {137}},\
  \bibinfo {pages} {136002} (\bibinfo {year} {2026})}\BibitemShut {NoStop}%
\bibitem [{\citenamefont {{See Supplementary Information}}()}]{suppl}%
  \BibitemOpen
  \bibfield  {author} {\bibinfo {author} {\bibnamefont {{See Supplementary
  Information}}},\ }\href@noop {} {}\BibitemShut {NoStop}%
\bibitem [{\citenamefont {Zhao}\ \emph {et~al.}(2012)\citenamefont {Zhao},
  \citenamefont {Lausberg}, \citenamefont {Kim}, \citenamefont {Tanatar},
  \citenamefont {Brando}, \citenamefont {Prozorov},\ and\ \citenamefont
  {Morosan}}]{zhao2012type}%
  \BibitemOpen
  \bibfield  {author} {\bibinfo {author} {\bibfnamefont {L.~L.}\ \bibnamefont
  {Zhao}}, \bibinfo {author} {\bibfnamefont {S.}~\bibnamefont {Lausberg}},
  \bibinfo {author} {\bibfnamefont {H.}~\bibnamefont {Kim}}, \bibinfo {author}
  {\bibfnamefont {M.~A.}\ \bibnamefont {Tanatar}}, \bibinfo {author}
  {\bibfnamefont {M.}~\bibnamefont {Brando}}, \bibinfo {author} {\bibfnamefont
  {R.}~\bibnamefont {Prozorov}}, \ and\ \bibinfo {author} {\bibfnamefont
  {E.}~\bibnamefont {Morosan}},\ }\href {\doibase 10.1103/PhysRevB.85.214526}
  {\bibfield  {journal} {\bibinfo  {journal} {Physical Review B}\ }\textbf
  {\bibinfo {volume} {85}},\ \bibinfo {pages} {214526} (\bibinfo {year}
  {2012})}\BibitemShut {NoStop}%
\bibitem [{\citenamefont
  {Onsager}(1931{\natexlab{a}})}]{onsager1931reciprocal1}%
  \BibitemOpen
  \bibfield  {author} {\bibinfo {author} {\bibfnamefont {L.}~\bibnamefont
  {Onsager}},\ }\href@noop {} {\bibfield  {journal} {\bibinfo  {journal}
  {Physical review}\ }\textbf {\bibinfo {volume} {37}},\ \bibinfo {pages} {405}
  (\bibinfo {year} {1931}{\natexlab{a}})}\BibitemShut {NoStop}%
\bibitem [{\citenamefont
  {Onsager}(1931{\natexlab{b}})}]{onsager1931reciprocal2}%
  \BibitemOpen
  \bibfield  {author} {\bibinfo {author} {\bibfnamefont {L.}~\bibnamefont
  {Onsager}},\ }\href@noop {} {\bibfield  {journal} {\bibinfo  {journal}
  {Physical review}\ }\textbf {\bibinfo {volume} {38}},\ \bibinfo {pages}
  {2265} (\bibinfo {year} {1931}{\natexlab{b}})}\BibitemShut {NoStop}%
\bibitem [{\citenamefont {Casimir}(1945)}]{casimir1945onsager}%
  \BibitemOpen
  \bibfield  {author} {\bibinfo {author} {\bibfnamefont {H.~B.~G.}\
  \bibnamefont {Casimir}},\ }\href@noop {} {\bibfield  {journal} {\bibinfo
  {journal} {Reviews of Modern Physics}\ }\textbf {\bibinfo {volume} {17}},\
  \bibinfo {pages} {343} (\bibinfo {year} {1945})}\BibitemShut {NoStop}%
\bibitem [{\citenamefont {Sato}\ \emph {et~al.}(1999)\citenamefont {Sato},
  \citenamefont {Kinokiri}, \citenamefont {Komatsubara},\ and\ \citenamefont
  {Harima}}]{sato1999effect}%
  \BibitemOpen
  \bibfield  {author} {\bibinfo {author} {\bibfnamefont {N.}~\bibnamefont
  {Sato}}, \bibinfo {author} {\bibfnamefont {T.}~\bibnamefont {Kinokiri}},
  \bibinfo {author} {\bibfnamefont {T.}~\bibnamefont {Komatsubara}}, \ and\
  \bibinfo {author} {\bibfnamefont {H.}~\bibnamefont {Harima}},\ }\href@noop {}
  {\bibfield  {journal} {\bibinfo  {journal} {Physical Review B}\ }\textbf
  {\bibinfo {volume} {59}},\ \bibinfo {pages} {4714} (\bibinfo {year}
  {1999})}\BibitemShut {NoStop}%
\bibitem [{\citenamefont {Gao}\ \emph {et~al.}(2026)\citenamefont {Gao},
  \citenamefont {Allen}, \citenamefont {Mustaf}, \citenamefont {Zhang},
  \citenamefont {Mishra}, \citenamefont {Lane}, \citenamefont {Zonno},
  \citenamefont {Gorovikov}, \citenamefont {Zhu}, \citenamefont {Yi},\ and\
  \citenamefont {Morosan}}]{gao2026quantum}%
  \BibitemOpen
  \bibfield  {author} {\bibinfo {author} {\bibfnamefont {Y.}~\bibnamefont
  {Gao}}, \bibinfo {author} {\bibfnamefont {K.}~\bibnamefont {Allen}}, \bibinfo
  {author} {\bibfnamefont {R.~A.}\ \bibnamefont {Mustaf}}, \bibinfo {author}
  {\bibfnamefont {Y.}~\bibnamefont {Zhang}}, \bibinfo {author} {\bibfnamefont
  {S.}~\bibnamefont {Mishra}}, \bibinfo {author} {\bibfnamefont
  {C.}~\bibnamefont {Lane}}, \bibinfo {author} {\bibfnamefont {M.}~\bibnamefont
  {Zonno}}, \bibinfo {author} {\bibfnamefont {S.}~\bibnamefont {Gorovikov}},
  \bibinfo {author} {\bibfnamefont {J.-X.}\ \bibnamefont {Zhu}}, \bibinfo
  {author} {\bibfnamefont {M.}~\bibnamefont {Yi}}, \ and\ \bibinfo {author}
  {\bibfnamefont {E.}~\bibnamefont {Morosan}},\ }\href {\doibase
  10.1103/8jjw-4r71} {\bibfield  {journal} {\bibinfo  {journal} {Physical
  Review B}\ }\textbf {\bibinfo {volume} {113}},\ \bibinfo {pages} {075115}
  (\bibinfo {year} {2026})}\BibitemShut {NoStop}%
\bibitem [{\citenamefont {Zhang}\ \emph {et~al.}(2005)\citenamefont {Zhang},
  \citenamefont {Tan}, \citenamefont {Stormer},\ and\ \citenamefont
  {Kim}}]{zhang2005experimental}%
  \BibitemOpen
  \bibfield  {author} {\bibinfo {author} {\bibfnamefont {Y.}~\bibnamefont
  {Zhang}}, \bibinfo {author} {\bibfnamefont {Y.-W.}\ \bibnamefont {Tan}},
  \bibinfo {author} {\bibfnamefont {H.~L.}\ \bibnamefont {Stormer}}, \ and\
  \bibinfo {author} {\bibfnamefont {P.}~\bibnamefont {Kim}},\ }\href@noop {}
  {\bibfield  {journal} {\bibinfo  {journal} {Nature}\ }\textbf {\bibinfo
  {volume} {438}},\ \bibinfo {pages} {201} (\bibinfo {year}
  {2005})}\BibitemShut {NoStop}%
\bibitem [{\citenamefont {Zhao}\ and\ \citenamefont
  {Wang}(2022)}]{zhao2022berry}%
  \BibitemOpen
  \bibfield  {author} {\bibinfo {author} {\bibfnamefont {W.}~\bibnamefont
  {Zhao}}\ and\ \bibinfo {author} {\bibfnamefont {X.}~\bibnamefont {Wang}},\
  }\href {\doibase 10.1080/23746149.2022.2064230} {\bibfield  {journal}
  {\bibinfo  {journal} {Advances in Physics: X}\ }\textbf {\bibinfo {volume}
  {7}},\ \bibinfo {pages} {2064230} (\bibinfo {year} {2022})}\BibitemShut
  {NoStop}%
\bibitem [{\citenamefont {Alexandradinata}\ and\ \citenamefont
  {Glazman}(2023)}]{alexandradinata2023fermiology}%
  \BibitemOpen
  \bibfield  {author} {\bibinfo {author} {\bibfnamefont {A.}~\bibnamefont
  {Alexandradinata}}\ and\ \bibinfo {author} {\bibfnamefont {L.}~\bibnamefont
  {Glazman}},\ }\href {\doibase 10.1146/annurev-conmatphys-040721-021331}
  {\bibfield  {journal} {\bibinfo  {journal} {Annual Review of Condensed Matter
  Physics}\ }\textbf {\bibinfo {volume} {14}},\ \bibinfo {pages} {261}
  (\bibinfo {year} {2023})}\BibitemShut {NoStop}%
\bibitem [{\citenamefont {Guo}\ \emph {et~al.}(2021)\citenamefont {Guo},
  \citenamefont {Alexandradinata}, \citenamefont {Putzke}, \citenamefont
  {Estry}, \citenamefont {Tu}, \citenamefont {Kumar}, \citenamefont {Fan},
  \citenamefont {Zhang}, \citenamefont {Wu} \emph
  {et~al.}}]{guo2021temperature}%
  \BibitemOpen
  \bibfield  {author} {\bibinfo {author} {\bibfnamefont {C.}~\bibnamefont
  {Guo}}, \bibinfo {author} {\bibfnamefont {A.}~\bibnamefont
  {Alexandradinata}}, \bibinfo {author} {\bibfnamefont {C.}~\bibnamefont
  {Putzke}}, \bibinfo {author} {\bibfnamefont {A.}~\bibnamefont {Estry}},
  \bibinfo {author} {\bibfnamefont {T.}~\bibnamefont {Tu}}, \bibinfo {author}
  {\bibfnamefont {N.}~\bibnamefont {Kumar}}, \bibinfo {author} {\bibfnamefont
  {F.-R.}\ \bibnamefont {Fan}}, \bibinfo {author} {\bibfnamefont
  {S.}~\bibnamefont {Zhang}}, \bibinfo {author} {\bibfnamefont
  {Q.}~\bibnamefont {Wu}},  \emph {et~al.},\ }\href {\doibase
  10.1038/s41467-021-26450-1} {\bibfield  {journal} {\bibinfo  {journal}
  {Nature Communications}\ }\textbf {\bibinfo {volume} {12}},\ \bibinfo {pages}
  {6213} (\bibinfo {year} {2021})}\BibitemShut {NoStop}%
\bibitem [{\citenamefont {Wu}\ \emph {et~al.}(2023)\citenamefont {Wu},
  \citenamefont {Hu}, \citenamefont {Graf}, \citenamefont {Liu}, \citenamefont
  {Deng}, \citenamefont {Fu}, \citenamefont {Kundu}, \citenamefont {Valla},
  \citenamefont {Petrovic},\ and\ \citenamefont {Wang}}]{wu2023coexistence}%
  \BibitemOpen
  \bibfield  {author} {\bibinfo {author} {\bibfnamefont {X.}~\bibnamefont
  {Wu}}, \bibinfo {author} {\bibfnamefont {Z.}~\bibnamefont {Hu}}, \bibinfo
  {author} {\bibfnamefont {D.}~\bibnamefont {Graf}}, \bibinfo {author}
  {\bibfnamefont {Y.}~\bibnamefont {Liu}}, \bibinfo {author} {\bibfnamefont
  {C.}~\bibnamefont {Deng}}, \bibinfo {author} {\bibfnamefont {H.}~\bibnamefont
  {Fu}}, \bibinfo {author} {\bibfnamefont {A.~K.}\ \bibnamefont {Kundu}},
  \bibinfo {author} {\bibfnamefont {T.}~\bibnamefont {Valla}}, \bibinfo
  {author} {\bibfnamefont {C.}~\bibnamefont {Petrovic}}, \ and\ \bibinfo
  {author} {\bibfnamefont {A.}~\bibnamefont {Wang}},\ }\href {\doibase
  10.1103/PhysRevB.108.245156} {\bibfield  {journal} {\bibinfo  {journal}
  {Physical Review B}\ }\textbf {\bibinfo {volume} {108}},\ \bibinfo {pages}
  {245156} (\bibinfo {year} {2023})}\BibitemShut {NoStop}%
\bibitem [{\citenamefont {Palacio}\ \emph {et~al.}(2023)\citenamefont
  {Palacio}, \citenamefont {Obando-Guevara}, \citenamefont {Chen},
  \citenamefont {Nair}, \citenamefont {Gonz{\'a}lez~Barrio}, \citenamefont
  {Papalazarou}, \citenamefont {Le~F{\`e}vre}, \citenamefont {Taleb-Ibrahimi},
  \citenamefont {Michel}, \citenamefont {Mascaraque},\ and\ \citenamefont
  {Tejeda}}]{palacio2023fermi}%
  \BibitemOpen
  \bibfield  {author} {\bibinfo {author} {\bibfnamefont {I.}~\bibnamefont
  {Palacio}}, \bibinfo {author} {\bibfnamefont {J.}~\bibnamefont
  {Obando-Guevara}}, \bibinfo {author} {\bibfnamefont {L.}~\bibnamefont
  {Chen}}, \bibinfo {author} {\bibfnamefont {M.~N.}\ \bibnamefont {Nair}},
  \bibinfo {author} {\bibfnamefont {M.~A.}\ \bibnamefont
  {Gonz{\'a}lez~Barrio}}, \bibinfo {author} {\bibfnamefont {E.}~\bibnamefont
  {Papalazarou}}, \bibinfo {author} {\bibfnamefont {P.}~\bibnamefont
  {Le~F{\`e}vre}}, \bibinfo {author} {\bibfnamefont {A.}~\bibnamefont
  {Taleb-Ibrahimi}}, \bibinfo {author} {\bibfnamefont {E.~G.}\ \bibnamefont
  {Michel}}, \bibinfo {author} {\bibfnamefont {A.}~\bibnamefont {Mascaraque}},
  \ and\ \bibinfo {author} {\bibfnamefont {A.}~\bibnamefont {Tejeda}},\ }\href
  {\doibase 10.1016/j.apsusc.2022.155477} {\bibfield  {journal} {\bibinfo
  {journal} {Applied Surface Science}\ }\textbf {\bibinfo {volume} {610}},\
  \bibinfo {pages} {155477} (\bibinfo {year} {2023})}\BibitemShut {NoStop}%
\bibitem [{\citenamefont {Ji}\ \emph {et~al.}(2016)\citenamefont {Ji},
  \citenamefont {Zhang}, \citenamefont {Fan}, \citenamefont {Li}, \citenamefont
  {Wang}, \citenamefont {Zhang}, \citenamefont {Plummer},\ and\ \citenamefont
  {Zhang}}]{ji2016giant}%
  \BibitemOpen
  \bibfield  {author} {\bibinfo {author} {\bibfnamefont {J.}~\bibnamefont
  {Ji}}, \bibinfo {author} {\bibfnamefont {A.}~\bibnamefont {Zhang}}, \bibinfo
  {author} {\bibfnamefont {J.}~\bibnamefont {Fan}}, \bibinfo {author}
  {\bibfnamefont {Y.}~\bibnamefont {Li}}, \bibinfo {author} {\bibfnamefont
  {X.}~\bibnamefont {Wang}}, \bibinfo {author} {\bibfnamefont {J.}~\bibnamefont
  {Zhang}}, \bibinfo {author} {\bibfnamefont {E.}~\bibnamefont {Plummer}}, \
  and\ \bibinfo {author} {\bibfnamefont {Q.}~\bibnamefont {Zhang}},\
  }\href@noop {} {\bibfield  {journal} {\bibinfo  {journal} {Proceedings of the
  National Academy of Sciences}\ }\textbf {\bibinfo {volume} {113}},\ \bibinfo
  {pages} {2349} (\bibinfo {year} {2016})}\BibitemShut {NoStop}%
\bibitem [{\citenamefont {Wulferding}\ \emph {et~al.}(2025)\citenamefont
  {Wulferding}, \citenamefont {Park}, \citenamefont {Tohyama}, \citenamefont
  {Park},\ and\ \citenamefont {Kim}}]{wulferding2025magnetic}%
  \BibitemOpen
  \bibfield  {author} {\bibinfo {author} {\bibfnamefont {D.}~\bibnamefont
  {Wulferding}}, \bibinfo {author} {\bibfnamefont {J.}~\bibnamefont {Park}},
  \bibinfo {author} {\bibfnamefont {T.}~\bibnamefont {Tohyama}}, \bibinfo
  {author} {\bibfnamefont {S.~R.}\ \bibnamefont {Park}}, \ and\ \bibinfo
  {author} {\bibfnamefont {C.}~\bibnamefont {Kim}},\ }\href@noop {} {\bibfield
  {journal} {\bibinfo  {journal} {Nature communications}\ }\textbf {\bibinfo
  {volume} {16}},\ \bibinfo {pages} {114} (\bibinfo {year} {2025})}\BibitemShut
  {NoStop}%
\bibitem [{\citenamefont {Yao}\ and\ \citenamefont
  {Yang}(2015)}]{yao2015topological}%
  \BibitemOpen
  \bibfield  {author} {\bibinfo {author} {\bibfnamefont {H.}~\bibnamefont
  {Yao}}\ and\ \bibinfo {author} {\bibfnamefont {F.}~\bibnamefont {Yang}},\
  }\href {\doibase 10.1103/PhysRevB.92.035132} {\bibfield  {journal} {\bibinfo
  {journal} {Phys. Rev. B}\ }\textbf {\bibinfo {volume} {92}},\ \bibinfo
  {pages} {035132} (\bibinfo {year} {2015})}\BibitemShut {NoStop}%
\bibitem [{\citenamefont {Meng}\ \emph {et~al.}(2015)\citenamefont {Meng},
  \citenamefont {Yang}, \citenamefont {Chen}, \citenamefont {Yao},\ and\
  \citenamefont {Kee}}]{meng2015evidence}%
  \BibitemOpen
  \bibfield  {author} {\bibinfo {author} {\bibfnamefont {Z.~Y.}\ \bibnamefont
  {Meng}}, \bibinfo {author} {\bibfnamefont {F.}~\bibnamefont {Yang}}, \bibinfo
  {author} {\bibfnamefont {K.-S.}\ \bibnamefont {Chen}}, \bibinfo {author}
  {\bibfnamefont {H.}~\bibnamefont {Yao}}, \ and\ \bibinfo {author}
  {\bibfnamefont {H.-Y.}\ \bibnamefont {Kee}},\ }\href {\doibase
  10.1103/PhysRevB.91.184509} {\bibfield  {journal} {\bibinfo  {journal} {Phys.
  Rev. B}\ }\textbf {\bibinfo {volume} {91}},\ \bibinfo {pages} {184509}
  (\bibinfo {year} {2015})}\BibitemShut {NoStop}%
\end{thebibliography}%

\end{document}